\documentclass[pre,aps,twocolumn,amsmath,amssymb]{revtex4-1}

\pdfoutput=1
\usepackage{hyperref}
\usepackage{graphicx}
\usepackage[usenames]{color}
\usepackage{bm}
\usepackage[final]{movie15}

\def\cal#1{\mathcal{#1}}
\def\eqq#1{Eq.~(\ref{#1})}
\def\eq#1{(\ref{#1})}
\def\av#1{\langle #1 \rangle}

\def\f#1{Fig.~\ref{#1}}

\def\c#1{~\cite{#1}}

\def\s#1{Section~\ref{#1}}
\def\kt{k_{\rm B}T}
\def\e{{\rm e}}

\def\beq{\begin{equation}}
\def\eeq{\end{equation}}
\def\bea{\begin{eqnarray}}
\def\eea{\end{eqnarray}}

\begin{document}

\title{Large deviations in the presence of cooperativity and slow dynamics}
\author{Stephen Whitelam}\email{{\tt swhitelam@lbl.gov}}
\affiliation{Molecular Foundry, Lawrence Berkeley National Laboratory, 1 Cyclotron Road, Berkeley, CA 94720, USA}
\begin{abstract}
We study simple models of intermittency, involving switching between two states, within the dynamical large-deviation formalism. Singularities appear in the formalism when switching is cooperative, or when its basic timescale diverges. In the first case the unbiased trajectory distribution undergoes a symmetry breaking, leading to a change of shape of the large-deviation rate function for a particular dynamical observable. In the second case the symmetry of the unbiased trajectory distribution remains unbroken. Comparison of these models suggests that singularities of the dynamical large-deviation formalism can signal the dynamical equivalent of an equilibrium phase transition, but do not necessarily do so.
\end{abstract}
\maketitle
  
\section{Introduction}
The dynamical behavior of models of statistical physics is described by trajectories, time-ordered sequences of configurations generated by a dynamical protocol\c{chandler1987introduction,newman1999monte}. The statistics of an ensemble of trajectories can be quantified by the dynamical large-deviation formalism\c{lebowitz1999gallavotti,touchette2011basic,ruelle2004thermodynamic,garrahan2007dynamical,Lecomte2007,Chetrite2014,touchette2009large,garrahan2009first,hedges2009dynamic,touchette2011large,speck2012large,vaikuntanathan2014dynamic,ray2017importance,horowitz2017stochastic,nyawo2017minimal}. This formalism allows the calculation of large-deviation rate functions for observables $A=aK$, such as entropy production\c{speck2012large} or configuration changes\c{garrahan2009first}, that are extensive in the length $K$ of a trajectory. For certain models, for long trajectories, the probability distribution of $a$ over the ensemble has a large-deviation form $\rho(a)\sim \e^{-X I(a)}$\c{evans2004rules,chetrite2013nonequilibrium,Chetrite2014,touchette2009large,garrahan2009first,maes2008canonical,giardina2011simulating,lecomte2007numerical,nemoto2014computation}. Here $X$ is the {\em rate} or {\em speed} of the large-deviation form, and $I(a)$ is the large-deviation {\em rate function} on this speed. These objects quantify the probability of observing a trajectory possessing a (potentially very unlikely) value of $a$. 

The emergence of singularities during the calculation of $I(a)$ can signal special behavior of the model under study. Usually $I(a)$ is calculated by Legendre transform of the scaled cumulant-generating function\c{lebowitz1999gallavotti}. Singularities of this function can appear for different reasons\c{baek2015singularities,touchette2009large}: they appear in the presence of long-tailed distributions, where the large-deviation principle does not apply\c{rebenshtok2014non,aghion2017large}, but also at special points in the parameter regimes of models for which the large-deviation principle does apply. Such models include kinetically constrained lattice models\c{garrahan2007dynamical,garrahan2009first,nemoto2017finite}, networks\c{horowitz2017stochastic,vaikuntanathan2014dynamic}, and models of driven\c{nyawo2017minimal,speck2012large} and active\c{pietzonka2016extreme} particles. Sometimes these singularities are described as {\em dynamical phase transitions}\c{garrahan2007dynamical,garrahan2009first,nemoto2017finite,horowitz2017stochastic,vaikuntanathan2014dynamic,nyawo2017minimal} and sometimes they are not\c{speck2012large,pietzonka2016extreme}. Regardless of the terminology used, it is clear that a variety of models, of a range of complexity, cause singularities in the large-deviation formalism. It is therefore natural to ask what range of model behaviors can give rise to such singularities, and how we can distinguish singularities on physical grounds.

 Here we study two models that for different reasons cause singularities in the large-deviation formalism. Both models involve switching between two states. In one case switching is cooperative, and depends on the history of the process (equivalent to a certain Markov process on an infinite state space). In the other case switching is non-cooperative. The models induce singularities in the large-deviation formalism for the distinct reasons sketched in \f{fig1}. For the cooperative model, singularities appear because the unbiased trajectory ensemble exhibits symmetry breaking -- i.e. the probability distribution of the dynamical observable becomes bimodal -- and the resulting large-deviation rate function becomes non-convex. This phenomenology is similar to that of magnetization in the equilibrium Ising model, in the sense that by varying a model parameter the probability distribution of the observable changes from unimodal to bimodal. By contrast, the non-cooperative model induces singularities in the large-deviation formalism when its switching time $\tau$ diverges, because the rate function on this speed becomes affine (linear).  However, the model supports only a single dynamical ``phase'', and this becomes clear when we view the large deviations of the model on the speed $K_\infty \equiv K/ \tau$: there $\rho(a)$ satisfies a large-deviation principle $\rho(a)\sim \e^{-K_\infty I_\infty(a)}$, where $I_\infty(a)$ is strictly convex with a unique minimum.
 \\~
 \indent In \s{main} we describe the models and their phenomenology. In \s{sing} we identify the distinct reasons they give rise to singularities of the large-deviation formalism. In \s{discuss} we compare the phenomenology of the two models to that of the equilibrium Ising model, and argue that singularities of the scaled cumulant-generating function do not necessarily indicate dynamical behavior akin to an equilibrium phase transition. We also comment on the value of using more than one method to reconstruct the large-deviation rate function. We conclude in \s{conc}.

\section{Models and Phenomenology} 
\label{main}
Consider the two-state models shown in \f{fig2}, evolved according to a discrete dynamics 
\\~
\beq
\label{me1}
{\bm P}(k+1) = {\bm W} {\bm P}(k).
\eeq
\\~
${\bm P}(k)$ is the column vector whose elements are $P(i,k)$, the probability of residing in state $i=0,1$ after $k$ steps of the dynamics, and ${\bm W}$ is the matrix whose $(j,i)^{\rm th}$ element is the transition probability $p(i \to j)$. Define a dynamical observable $A$\c{garrahan2009first} that is increased by 0 or 1 for each transition made into state 0 or 1, respectively; $a=A/K$ is the intensive counterpart of $A$ for a trajectory of $K$ steps. We are interested in the probability distribution $\rho(a)$ for large $K$, and whether it adopts the large-deviation form $\rho(a) \sim \e^{-K I(a)}$. In Model 1 [\f{fig2}(a)], which is a re-writing of the irreversible growth model of Refs.\c{klymko2016similarity,klymko2017rare}, the transition probabilities depend on $a$~\footnote{See Refs.\c{harris2009current,maes2009dynamical} for additional examples of systems whose rates possess memory.}: the quantity $\gamma(a)$ is 
\beq
 \gamma(a) = \frac{\e^{-Jm(a)}}{2 \cosh J m(a)},
 \eeq
where $m(a) \equiv 2 a -1$ and $J$ is a parameter. The trajectories of this model undergo, in the ``thermodynamic'' (i.e. large-$K$) limit, a continuous dynamical (i.e. nonequilibrium) phase transition as $J$ is varied~\footnote{The reversible model of those references, which can be recast as a 4-state model with history-dependent rates, also displays a first-order nonequilibrium phase transition.}. \f{fig3}(a) illustrates this phase transition. The typical activity $a_0$ is governed by the equation $m_0 = \tanh (J m_0)$, with $a_0 = (m_0+1)/2$. For $J\leq1$ the typical activity is $a_0=1/2$, while for $J>1$ there are two typical values $a_0^\pm$ of the activity, symmetrically disposed about the value $1/2$. Here, trajectories of two distinct types coexist. The point $J=1$ is a nonequilibrium critical point, where the unbiased trajectory-ensemble susceptibility, $\chi = K (\av{a^2}-\av{a}^2)$, grows as $K^{2/3}$, and so diverges in the large-$K$ limit (the bracket $\av{\cdot}$ denotes an average over unbiased trajectories of the model).
 \begin{figure}[] 
    \centering
    \includegraphics[width=\linewidth]{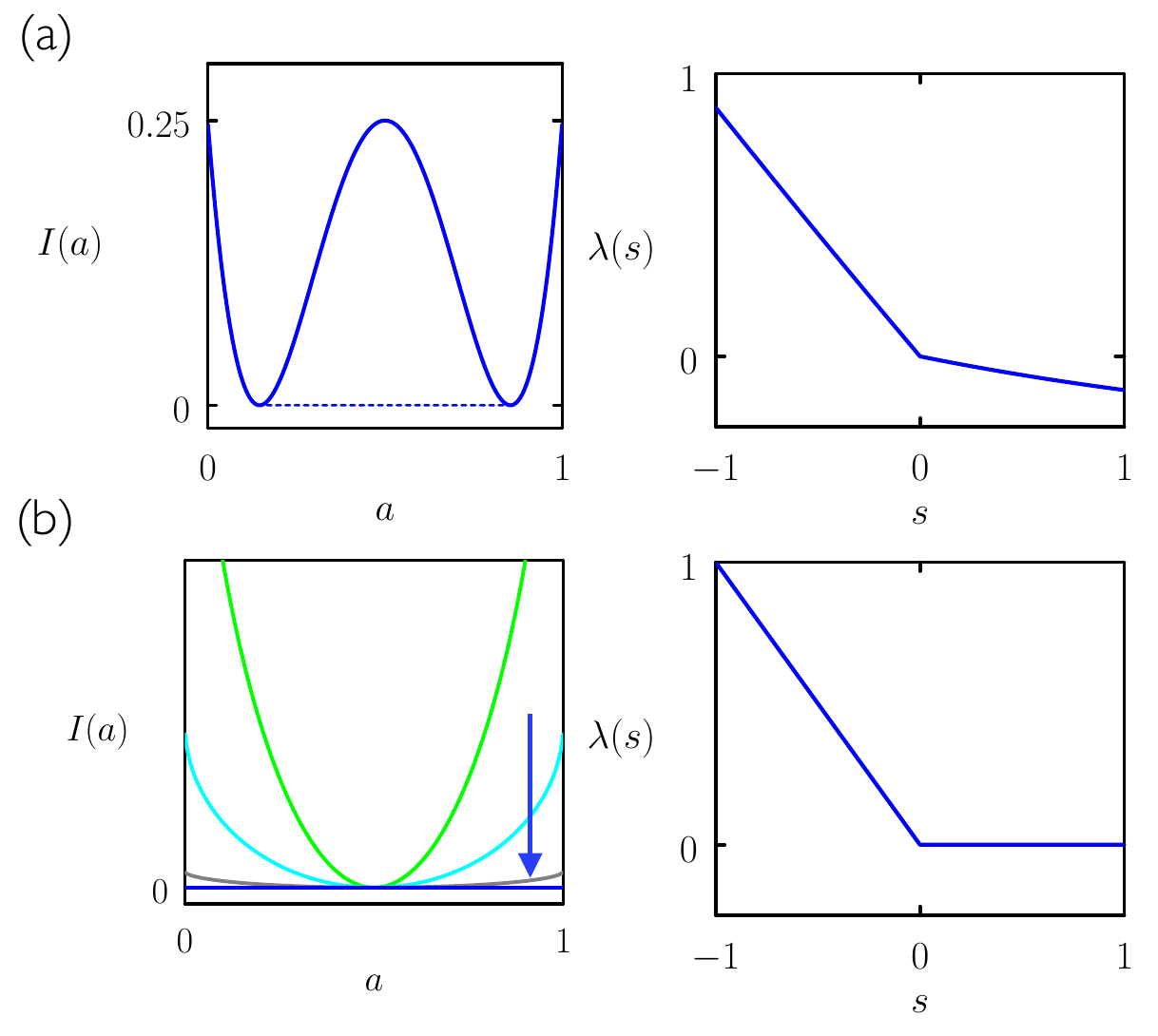} 
    \caption{Distinct reasons for singularities of the dynamical large-deviation formalism exhibited by models considered here (see also Ref.\c{touchette2009large}). (a) Symmetry breaking of the unbiased trajectory distribution results in two distinct stable dynamical phases, and the rate function $I(a)$ describing this behavior is non-convex. The Legendre transform $\lambda(s)$ of such a function is singular; the dashed line is the Legendre transform of $\lambda(s)$. (b) The rate function of the model becomes affine in response to a growing timescale, inducing a singularity in $\lambda(s)$. In this case there is no broken symmetry of the trajectory ensemble: when viewed on a more natural speed, one accounting for the large timescale, the large-deviation rate function remains strictly convex with a unique minimum.}
    \label{fig1}
 \end{figure} 
   \begin{figure}[b] 
    \centering
    \includegraphics[width=\linewidth]{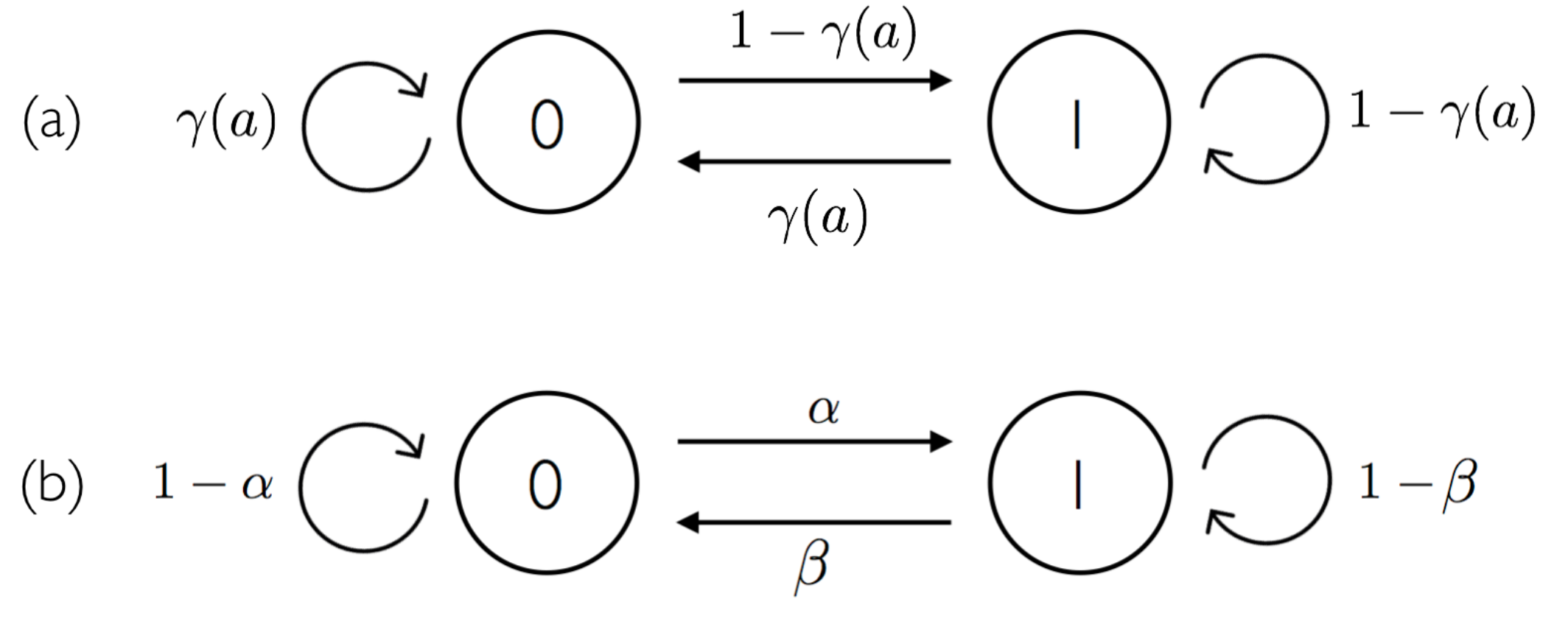} 
    \caption{Two-state models with transition probabilities as indicated. We increase the dynamical order parameter $A$ by an amount $0$ (resp. $1$) upon any transition into state 0 (resp. 1). (a) In Model 1, the transition probabilities depend upon $a=A/K$, where $K$ is the number of steps of the dynamics. (b) In Model 2, transition probabilities are fixed.}
    \label{fig2}
 \end{figure} 
 
In Model 2 [\f{fig2}(b)], the transition probabilities are fixed (see example IV.4 of\c{touchette2009large}). This model can display large fluctuations, but exhibits no symmetry breaking. For small values of the rate constants $\alpha,\beta$, the dynamics involves runs of states 0 or 1, the lengths of which are exponentially distributed with mean $\alpha^{-1}$ or $\beta^{-1}$. Runs therefore have finite length, and sufficiently long trajectories must exhibit a value $a_0=1/2$ of the dynamical observable $a$; see \f{fig3}(b). This model could be considered to be a minimal representation of a one-dimensional active walker\c{pietzonka2016extreme}, whose displacement after $K$ steps is $(2a -1)K$.
 
\section{Singularities of the dynamical large-deviation formalism}
\label{sing}
 
 \subsection{Model 1 -- cooperative dynamics}

Models 1 and 2 behave differently, but each can induce singularities in the dynamical large-deviation formalism. In \f{fig4}(a) we show large-deviation rate functions on speed $K$, $I(a)= -\lim_{K \to \infty} K^{-1} \ln \rho(a)$, for Model 1 (gray lines) for various values of $J$, calculated using the reference-model method of Refs\c{klymko2017rare,lattice2} (a form of nonequilibrium umbrella sampling). Rate functions have the form
\bea
 \label{eden}
I(m(a))&=&  \frac{1-m}{2} \ln \left(1-m\right)+\frac{1+m}{2}\ln \left( 1+m \right)\nonumber \\&-& J m^2 + \ln \cosh Jm,
\eea
  
which shows a change of shape as $J$ passes through the critical value 1. This change of shape reflects the symmetry breaking of the trajectory ensemble shown in \f{fig3}(a). For $J<1$ the rate function has a unique minimum, at which it is locally quadratic, i.e. $I(a) \propto (a-1/2)^2$. For $J=1$ the rate function is non-quadratic about its minimum ($I(a) \propto (a-1/2)^6$), which is the feature that ensures critical fluctuations and a diverging unbiased trajectory susceptibility $K (\av{a^2}-\av{a}^2)$\c{ellis1985large}. For $J>1$ the rate function is non-convex, reflecting coexistence of different types of trajectory: the zeros of a rate function indicate typical behavior\c{touchette2009large}, and in this regime we have two types of typical behavior. 
    \begin{figure*}[t] 
    \centering
    \includegraphics[width=0.9\linewidth]{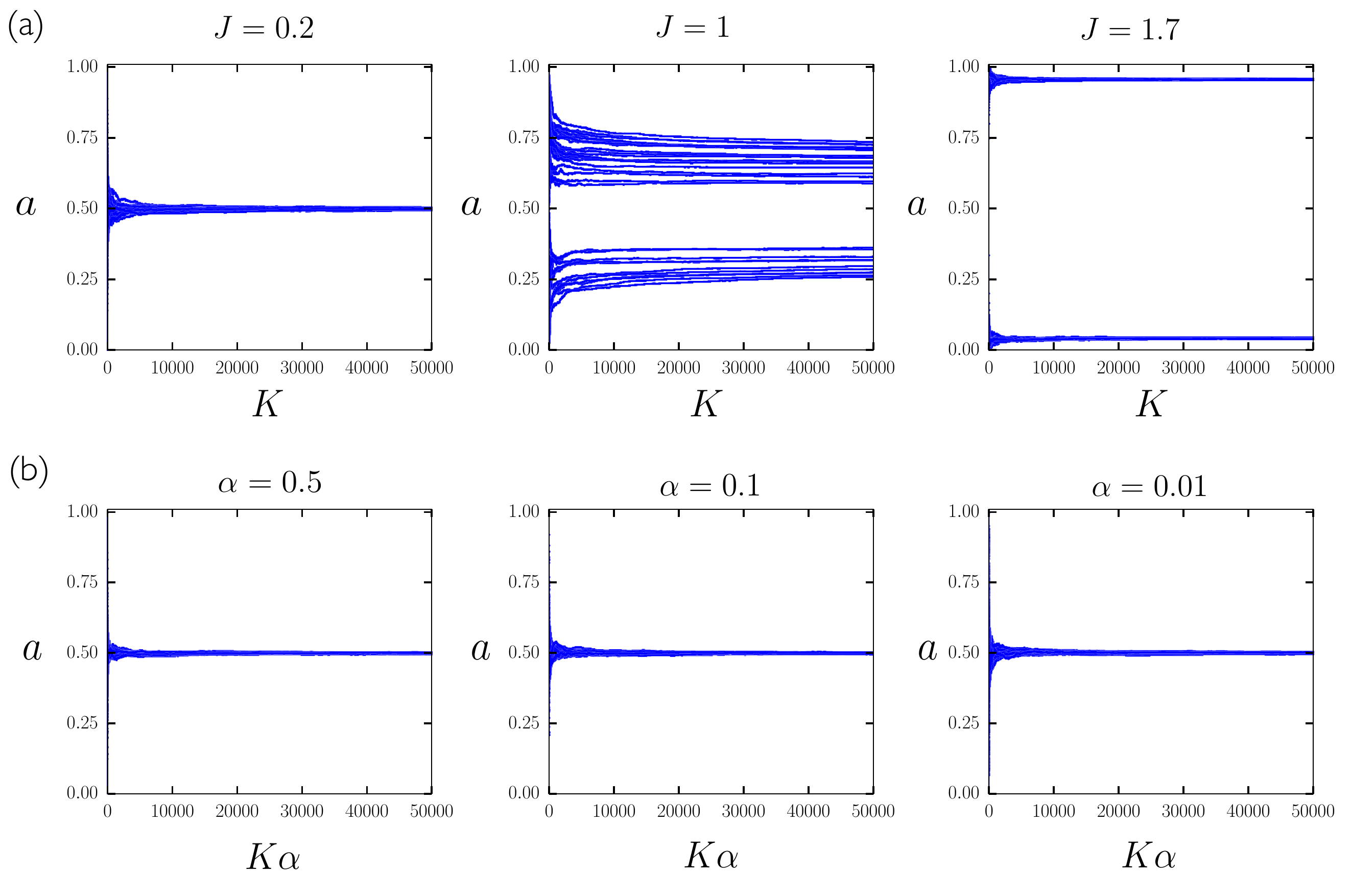} 
    \caption{Ensembles of trajectories (time-integrated intensive activity $a=A/K$ versus step number $K$) for the models of \f{fig2}. (a) For Model 1, the ensemble changes from having one typical value to two typical values as we pass through the nonequilibrium phase transition at $J=1$; there, trajectory fluctuations are critical in the sense that the susceptibility $K (\av{a^2}-\av{a}^2)$ diverges with trajectory length $K$. The large-deviation rate functions describing this phase transition are shown in \f{fig4}. (b) For Model 2 ($\alpha=\beta$), little difference is seen as the switching rate varies from fast to slow, if the measure of time is scaled by the dominant timescale (the horizontal axis is $K/\alpha^{-1}$). This behavior reflects the fact that the large-deviation rate function describing the trajectory ensemble changes scale but not shape as $\alpha^{-1}$ diverges (see the forms $I_0$ and $I_\infty$ in \f{fig4}).}
    \label{fig3}
 \end{figure*} 
 \begin{figure*}[] 
    \centering
    \includegraphics[width=\linewidth]{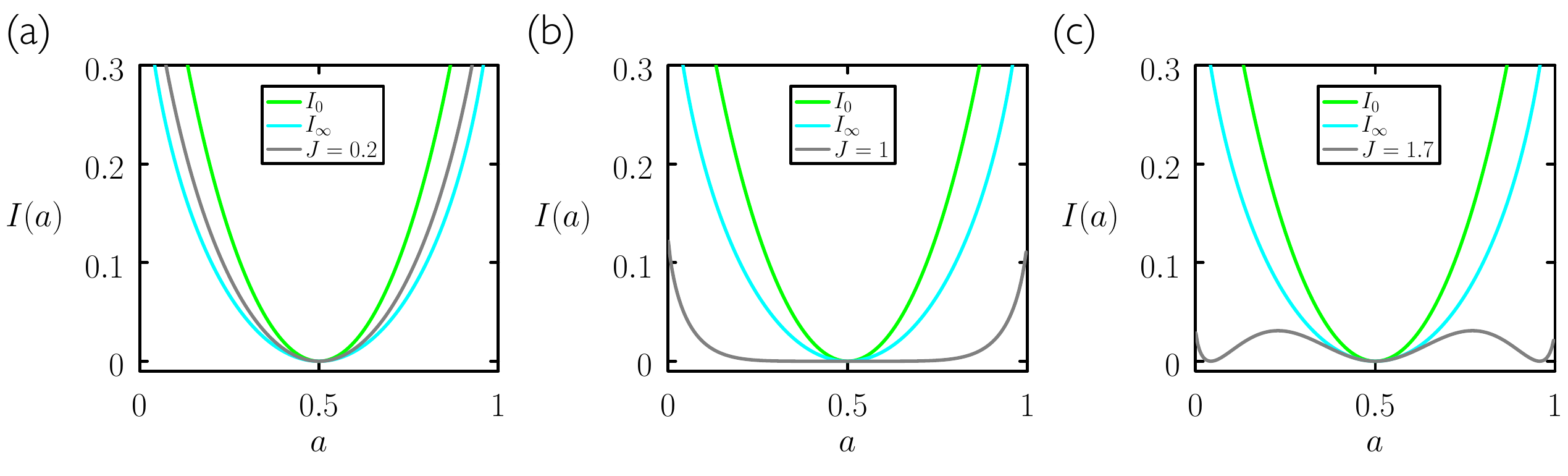} 
    \caption{Large-deviation rate functions for Model 1 (gray lines), illustrating the change of shape of the function as we pass through the nonequilibrium phase transition at $J=1$. The corresponding symmetry-breaking of the trajectory ensemble is shown in \f{fig3}(a). The forms $I_0$ and $I_\infty$ are short-timescale and long-timescale rate functions for Model 2; these are strictly convex with a unique minimum. Corresponding trajectories are shown in \f{fig3}(b).}
    \label{fig4}
 \end{figure*} 
 ~
 \subsection{Model 2 -- slow dynamics}
  \label{mod2}
  
In \f{fig4} (colored lines) we show the shape of the rate function, for Model 2, in the limits of slow and fast switching. These functions have similar shape, being quadratic about their unique minimum $a_0=1/2$. In this model the scale of the rate function changes as we vary switching time, but its shape does not, and there is no symmetry breaking of $\rho(a)$: see \f{fig3}(b). 

These rate functions can be calculated by application of the standard large-deviation formalism\c{lebowitz1999gallavotti,dembo2010large,touchette2009large,garrahan2009first}. Such methods use the non-probability-conserving dynamics  
\beq
{\bm P}_s(k+1) = {\bm W}_s {\bm P}_s(k),
\eeq
whose generator ${\bm W}_s$ is the matrix whose $(j,i)^{\rm th}$ element is $p_s(i \to j) =\e^{-s \sigma(i \to j)} p(i \to j)$. Here $\sigma(i \to j)$ is the change of $A$ upon moving from state $i$ to $j$. For dynamical systems the path ensemble generated by this formalism is sometimes called the ``$s$-ensemble''\c{garrahan2009first}. For Model 2,
\bea
\label{tilt}
{\bm W}_s  = \left(
\begin{array}{cc}
 1-\alpha & \beta  \\
 \alpha  \e^{-s} & (1-\beta ) \e^{-s}
\end{array}
\right).
\eea
If $\xi$ is the principal eigenvalue of ${\bm W}_s$, then $\lambda(s) = \ln \xi$ is the scaled cumulant-generating function, which reads
\begin{widetext}
\beq
\label{ev1}
\lambda(s)=\ln \left[\frac{1}{2} \e^{-s} \left(\sqrt{\left((\alpha -1)
   \e^s+\beta -1\right)^2+4 \e^s (\alpha +\beta -1)}+1-\beta +\e^s (1-\alpha)\right)\right].
\eeq
\end{widetext}
Legendre transform of $\lambda(s)$ yields a rate function
 \beq
\label{leg}
I_{\rm c}(a) = \max_s(-sa-\lambda(s)),
\eeq
which is equal to the true rate function $I(a)$ when the latter is convex; if not, $I_{\rm c}(a)$ returns the convex hull of the true rate function\c{dembo2010large,dinwoodie1993identifying,touchette2009large,chetrite2013nonequilibrium,Chetrite2014}.
\begin{figure*}[t] 
    \centering
    \includegraphics[width=\linewidth]{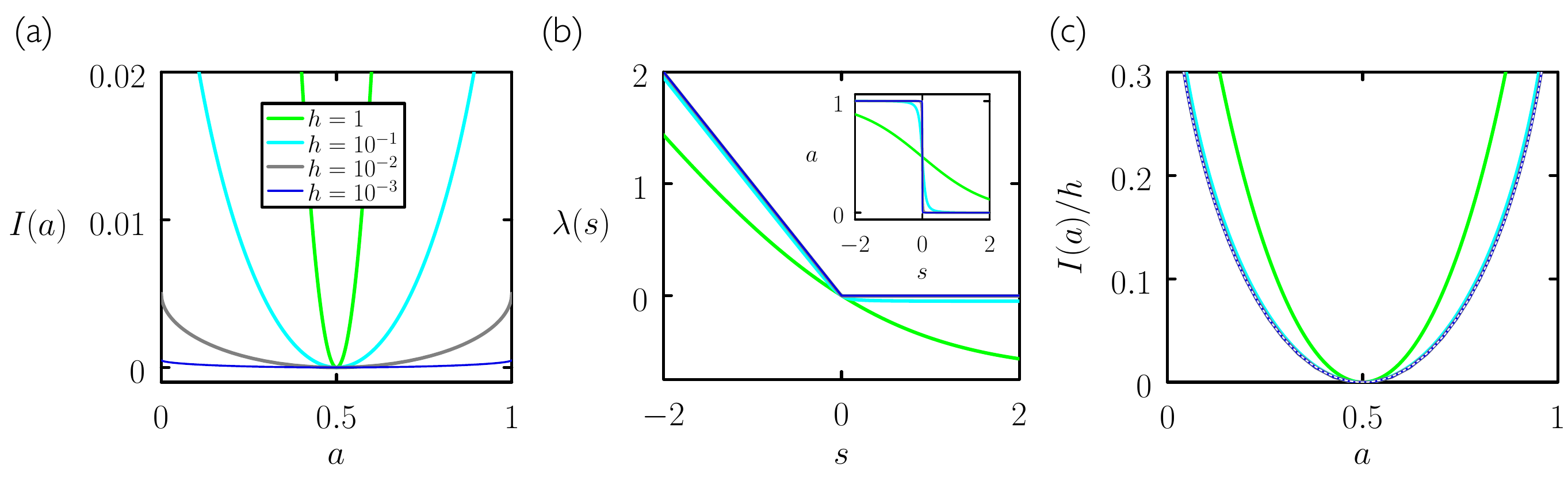} 
    \caption{(a) Large-deviation rate function $I(a)=-\lim_{K \to \infty} K^{-1} \ln \rho(a)$ for Model 2 of \f{fig1} (parameters $\alpha=\beta=h/2$), computed using the dynamical large-deviation formalism. Panel (b) shows the emerging singularities of $\lambda(s)$ and $a(s)=-\partial_s \lambda$ in the limit of slow switching. (c) Rate functions rescaled by $h$. For $h \to 0$, $I(a)/h$ adopts the limiting form \eq{limit} (white dashed line). This collapse reveals that intermittency in the limit of diverging switching time can be described by a large-deviation principle $\rho(a) \sim \e^{-K_\infty I_\infty(a)}$, in which the speed $K_\infty = K/ h^{-1}$ depends on the trajectory length $K$ {\em and} the divergent timescale $h^{-1}$.}
    \label{fig5}
 \end{figure*} 
 
The rate function and scaled cumulant-generating function obtained in this way are shown in \f{fig5}(a,b) for the case $\alpha=\beta=h/2$. As $h$ becomes small, the rate function becomes broad, indicating long-timescale intermittency and large fluctuations; similar features are seen in continuous-space models of active walkers\c{pietzonka2016extreme}. The scaled cumulant-generating function becomes increasingly sharp, and in the limit $h \to 0$ becomes singular, i.e.
\beq
\lambda(s) = \left\{\begin{array}{cc}
 0 & (s\geq0)\\
-s & (s <0).
\end{array} \right. 
\eeq
The typical activity $a(s) = -\partial_s \lambda(s)$ of the biased ensemble is shown in the inset of \f{fig5}(b). This switches increasingly sharply, with $s$, between two types of behavior as $h$ is made small, and in the limit $h \to 0$ jumps discontinuously with $s$. For general $\alpha, \beta$ the susceptibility of the {\em biased} ensemble, $\chi_s = - \partial_s^2 \lambda(s)$, takes its maximum value 
\beq
\chi_s^\star=\sqrt{\frac{(1-\alpha)(1-\beta)}{16 \alpha \beta}}
\eeq
at the point $s^\star=\ln [(\beta-1)/(\alpha-1)]$, and so diverges as $\sim (\alpha \beta)^{-1/2}$ as the rate constants vanish.
  \begin{figure*}[t] 
    \centering
    \includegraphics[width=\linewidth]{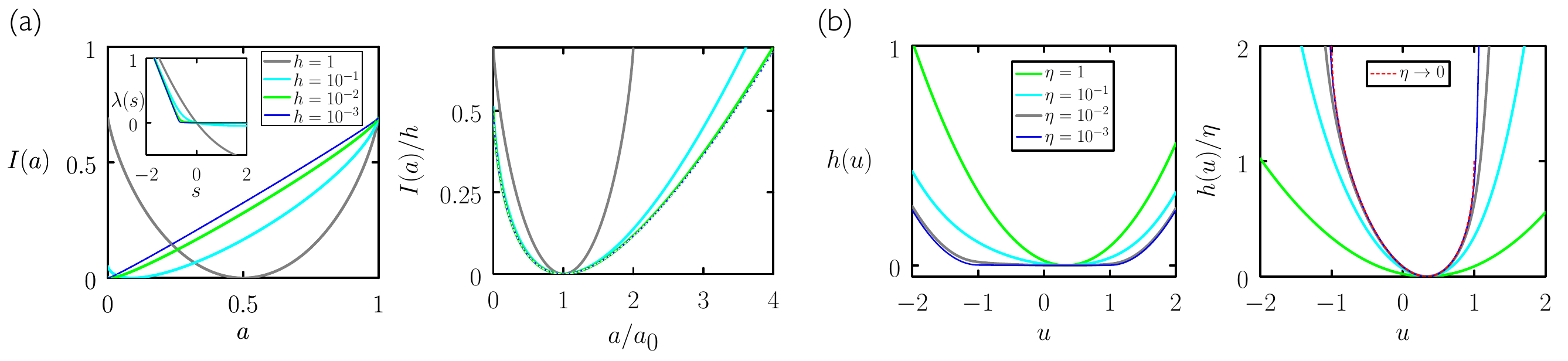} 
    \caption{(a) As \f{fig5}, for the parameter choice $\beta =1/2$ and $\alpha = h/2$. As $h$ becomes small, $I(a)=-\lim_{K \to \infty} K^{-1} \ln \rho(a)$ acquires a linear form because of the resulting separation of timescales. As a consequence, $\lambda$ changes rapidly with $s$ (inset), eventually becoming singular. On timescales of observation much larger than $h^{-1}$, however, a well-defined rate function emerges: here $\rho(a) \sim \e^{-K_\infty I_\infty(a/a_0)}$, where $K_\infty =K/h^{-1}$ and the function $I_\infty(a/a_0)$ is given by \eqq{eqs1} (white dashed lines). (b) Rate functions and rescaled rate functions for the active walker of Ref.\c{pietzonka2016extreme} (Fig. 1, left column of that reference) indicate that a large-deviation principle emerges in the limit of slow dynamics ($\eta \to 0$) if the speed of the large-deviation form contains the large timescale. The solid lines are obtained by numerical evaluation of \eqq{a1}; the red dashed line is \eqq{a3}.}
    \label{fig6}
 \end{figure*}
 
Although singularities of the large-deviation formalism emerge in Model 2 in the limit of slow switching, there is no symmetry breaking of the underlying (unbiased) trajectory ensemble: there is a single stable dynamical phase, specified by $a_0 = 1/2$. Singularities result instead from the fact that the switching time has exceeded our observation time, causing the rate function on speed $K$ to become linear. When the basic unit of time, $K$, is rescaled by the mean switching time $\alpha^{-1}$, the trajectory ensemble becomes almost independent of $\alpha$ [see \f{fig3}(b)]. Correspondingly, rate functions collapse to a limiting form $I_\infty(a)$ when the speed of the large-deviation form is scaled by the diverging switching timescale: see \f{fig5}(c). 

 Thus the switching process for the parameter choice $\alpha=\beta=h/2$ is described, in the diverging-timescale limit, by the behavior $-K^{-1} \ln \rho(a) \sim I(a) \sim I_\infty(a) h$. We can rewrite this in the form of a large-deviation principle $\rho(a) \sim \e^{-K_\infty I_\infty(a)}$ with speed $K_\infty=K/h^{-1}$. (A related scaling is seen in the low-noise limit of diffusion processes\c{PhysRevE.94.032101}.) The new speed depends on the trajectory length $K$ {\em and} the divergent timescale $h^{-1}$. The physics of the process therefore becomes insensitive to $h$, for sufficiently small values of the parameter, but we can discern this only if our observation time grows faster than the switching time. 
 
The rate function $I_\infty(a)$ is strictly convex, and so is Legendre dual to a smooth function, i.e. the singular behavior of $\lambda(s)$ can be cured by replacing $s \to sh$ (so effecting a rescaling of time) before taking the limit $h \to 0$. Maclaurin expansion in $h$ yields $\lambda(s h ) = h \lambda_\infty(s) + {\cal O}(h^2)$, where
\beq
\lambda_\infty(s) = \frac{1}{2} \left(\sqrt{s^2+1}-s-1\right)
\eeq
 is analytic in $s$ at $s=0$. The time-rescaled version of \eqq{leg} is then
\bea
\label{limit}
I_\infty(a)=\lim_{h \to 0} I_{\rm c}(a)/h &=& \max_s(-s a - \lambda_\infty(s)) \nonumber \\
&=& \frac{1}{2}-\sqrt{(1-a) a},
\eea
shown as a white dashed line in \f{fig5}(c). In this rescaled reference frame we have, to leading order, Gaussian fluctuations about the mean $a_0=1/2$ with variance of order unity, i.e. $I_\infty(a) \approx \left(a-a_0\right)^2$.
 
In \f{fig6}(a) we show that similar behavior obtains if there exists in the model a separation of timescales $\alpha^{-1} \gg \beta^{-1}$, in the limit $\alpha \to 0$. The rate function acquires a linear form, indicating exponential decay (with rate constant $\propto \beta$) to state 0. As a consequence, the quantity $\lambda(s)$ shows sharp features (inset), becoming singular in the limit $\alpha \to 0$. But there is no symmetry breaking of the unbiased trajectory ensemble, and upon suitable rescaling [panel (b)] we see that the switching process supports only one dynamical phase [the white dashed line is the form
 \beq
 \label{eqs1}
I_\infty(v) =  \frac{1}{4} (3-w)+v \ln \left(\frac{8 v}{4 v+w+1}\right),
 \eeq 
 in which $v \equiv a/a_0$, $a_0=h/(1+h)$, and $ w \equiv \sqrt{8 v+1}$].
 
 \section{Discussion}
 \label{discuss}
 
Model 1 and 2 display singularities of the dynamical large-deviation formalism. A growing body of work shows that similar singularities arise in a range of dynamical models, of a range of complexity, and sometimes these singularities are interpreted as dynamical phase transitions\c{garrahan2007dynamical,garrahan2009first,nemoto2017finite,horowitz2017stochastic,vaikuntanathan2014dynamic,nyawo2017minimal}. Analysis of the present models, which display distinct phenomenology, suggests that singularities of the dynamical large-deviation formalism can signal the dynamical equivalent of an equilibrium phase transition, but do not necessarily do so. To make this point more explicit, consider a familiar example of a model possessing an equilibrium phase transition, the 2D Ising model in the canonical ensemble (assume coupling $J/T$ and zero magnetic field). A detailed large-deviation analysis of the 2D Ising model is a challenging task\c{pfister1997large,ioffe1995exact,ioffe1994large}, but we can extract some of the essential ideas from idealized limits of the model. For $T$ far above the critical temperature we approximate the model as a collection of $N$ noninteracting spins. In equilibrium each spin is then up or down with equal likelihood, and the probability distribution of the intensive magnetization $m = M/N$ is that of the unbiased coin toss with $N$ trials, i.e. $P(m) \sim \e^{-N I_N(m)}$, with $I_N(m) = \frac{1}{2}(1-m) \ln(1-m) +\frac{1}{2}(1+m) \ln(1+m)$\c{varadhan2010large}. This function, which is convex with a unique minimum at $m=0$, is plotted in \f{fig_ising}(a).
 \begin{figure*}[htbp] 
    \centering
    \includegraphics[width=\linewidth]{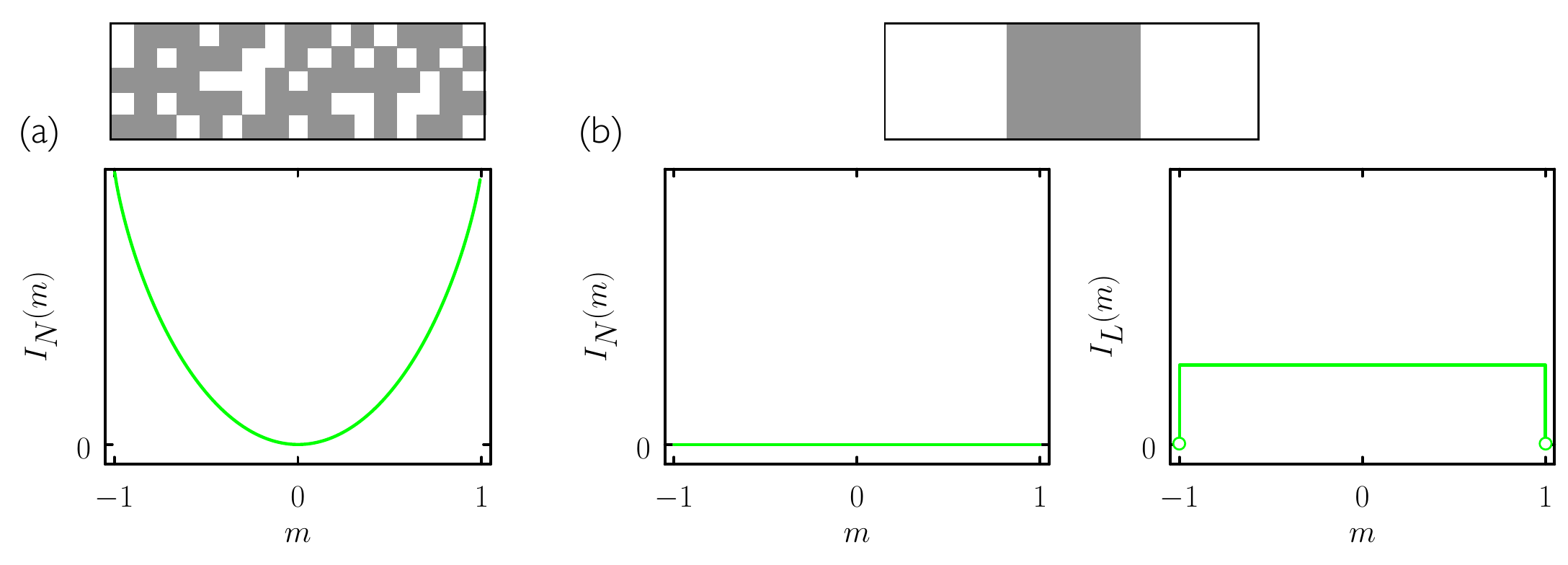} 
    \caption{A caricature of the probability distribution function of magnetization of the equilibrium 2D Ising model far above (a) and far below (b) the critical temperature $T_c$. The function $I_X(m) \equiv -\lim_{X \to \infty} X^{-1} \ln P(m)$. (a) Above $T_c$ the speed of the large-deviation principle is system size $N$, and the rate function is strictly convex. (b) Below $T_c$ the rate function on speed $N$ is affine (left), but on the surface speed $L$ (right) is non-convex. The two minima of this latter function describe two stable phases. Models 1 and 2, which both exhibit singularities of the dynamical large-deviation formalism, display the dynamical analog of some of the Ising model's behaviors, but to different extents.}
    \label{fig_ising}
 \end{figure*}
   \begin{figure*}[] 
    \centering
    \includegraphics[width=\linewidth]{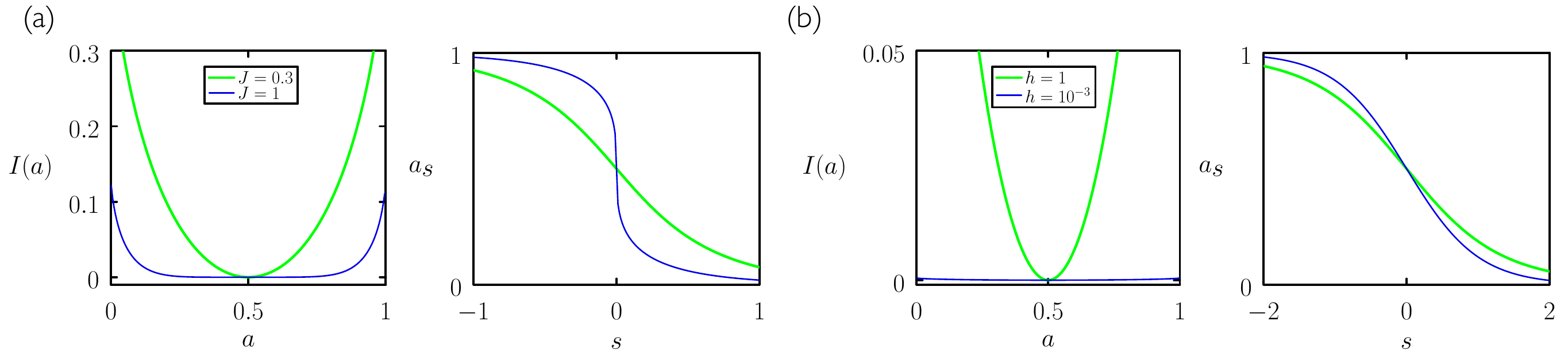} 
    \caption{For Model 1 (a) and Model 2 (b), we show rate functions $I(a)$ computed using the reference-model method of Refs.\c{klymko2017rare,lattice2} (a form of nonequilibrium umbrella sampling), together with the typical activity $a_s$ of the associated biased ensemble. $a_s$ changes sharply with $s$ at the nonequilibrium phase transition of Model 1, but changes smoothly in the limit of slow dynamics of Model 2 (cf. the dynamical large-deviation formalism, \f{fig5}(b)).}
    \label{fig7}
 \end{figure*}
 \\~
\indent Below the critical temperature the phenomenology of the model changes, because it can support a stable interface between up- and down-spin phases. In the limit of low temperature we can consider an idealized set of configurations of the kind shown in \f{fig_ising}(b). An up-spin region of width $W \in \{0,1,\dots,L\}$ is separated by two planar interfaces from a region of down spins, so that $m=2W/L-1$. The box has periodic boundaries and is of dimensions $L$ (horizontal) and $aL<L$ (vertical); the interface runs in the short direction. The relative energies of configurations are
\begin{equation}
  E(m)=\begin{cases}
    0 & (m=1)\\
   4 J a L& (-1<m<1)\\
     0 &(m=-1).
 \end{cases}
 \end{equation}
 The probabilities of configurations in thermal equilibrium are then
 \begin{equation}
 \label{pm}
  P(m)=Z^{-1} \times \begin{cases}
    1 & (m=1)\\
    L \e^{-\kappa L}& (-1<m<1)\\
      1 &(m=-1),
 \end{cases}
 \end{equation}
 where $\kappa \equiv 4 J a/(\kt)$ and $Z = 2 + L(L-1) \e^{-\kappa L}$ is the canonical partition function. The factor of $L$ in the middle line of \eq{pm} comes from the fact that the strip of up spins can be placed at any of $L$ points along the long direction of the box. 
 
From \eq{pm} we see that the large-deviation rate function on the bulk speed $N$ is zero everywhere, i.e. $I_N(m) \equiv -\lim_{N \to \infty} N^{-1} \ln P(m) = 0$ for all $m$. The large-deviation principle operates instead on the surface speed\c{touchette2009large} $L$, i.e.
  \begin{equation}
  I_L(m) = \begin{cases}
    0 & (m=1)\\
   \kappa& (-1<m<1)\\
      0 &(m=-1),
 \end{cases}
 \end{equation}
  with $I_L(m) \equiv -\lim_{L \to \infty} L^{-1} \ln P(m)$. As shown in \f{fig_ising}(b), the rate function on this speed is non-convex, with zeros at the two typical (equilibrium) values $m = \pm 1$. This rate function implies a bimodal probability distribution of magnetization whose peaks become sharper as the box gets bigger. In the Ising model we see a change between a one-phase region at high $T$, and a two-phase region at low $T$, with associated changes in the speed of the large-deviation principle and the shape of the rate function. Application of a magnetic field selects one of these phases over the other. Singularities are associated with these behaviors: in the two-phase region the Helmholtz free energy develops a kink, and the Gibbs free energy, its Legendre transform, develops an affine portion\c{binney1992theory}. At coexistence the probability distribution of magnetization is bimodal, controlled by the surface free energy $\propto I_L(m)$\c{binney1992theory,chandler1987introduction}.
  
Models 1 and 2 both display large-deviation singularities, but display the dynamical analog of Ising model behavior to differing degrees. Model 1 is similar to the Ising model in the sense that its rate function changes from being convex in one region of parameter space, describing a single typical behavior, to being non-convex in a different region of parameter space, describing multiple typical (coexisting) behaviors [\f{fig3}(a) \& \f{fig4}]. Between these two regions is a critical point at which the probability distribution exhibits anomalous fluctuations [\f{fig3}(a) \& \f{fig4}(b)]. In this sense it is natural to consider Model 1 to support a phase transition. Model 1 is unlike the Ising model in that the speed of its large-deviation principle does not change at the transition point, remaining $K$ (in this respect it is more similar to the mean-field version of the Ising model\c{paga2017large}). 
 
Model 2 is similar to the Ising model in that the speed of its large-deviation principle changes in a certain regime of parameter space, but it is different in that it only supports one stable phase [\f{fig3}(b) \& \f{fig4}]. In Model 2, the emergence of an affine rate function on speed $K$ [\f{fig5}(a), \f{fig6}(a)] in the limit of vanishing rate constants looks similar to the affine rate function $I_N(m)$ of the Ising model in the two-phase region on the bulk speed \f{fig_ising}(b). However, the large-deviation principle for the two models is more naturally formulated on the speeds $Kh$ and $L$, and on those speeds the rate functions are qualitatively different: Model 2 supports only one phase, while the Ising model supports coexisting phases [compare \f{fig5}(c) \& \f{fig6}(a) with $I_L(m)$ in \f{fig_ising}(b)]. 
\\~
 \indent Differences between Model 1- and Model 2 types of large-deviation singularities can also be identified by using different methods of reconstructing the rate function. The standard method of doing so is to compute the scaled cumulant-generating function and take its Legendre transform\c{lebowitz1999gallavotti, dembo2010large,touchette2009large}. However, this procedure cannot be used to reconstruct non-convex functions\c{dinwoodie1993identifying,ioffe1993two,touchette2009large}, and gives rise to singularities if the rate function is affine (linear) or non-convex. This method provides the right-hand panels of \f{fig1}, from which we could not tell if the underlying rate function corresponds to either of the scenarios shown in the left-hand panels. For the scenario shown in \f{fig1}(b) we then have an apparent contradiction. The biased process corresponding to the scaled cumulant-generating function displays a typical value of its order parameter that is given by the derivative of the function\c{touchette2009large}. This typical value therefore jumps discontinuously when the scaled cumulant-generating function is kinked. This is the case for Model 2, for instance: see \f{fig5}(b), \s{app1}, and \f{fig_supp}(b). Thus a model whose unbiased dynamics supports only one dynamical phase gives rise to phase transition-like phenomenology in a biased ensemble. This difference presents an apparent inconsistency. The object controlling the behavior of the biased ensemble, which shows phase transition-like phenomenology, is the scaled cumulant-generating function. But this function is entirely determined (via Legendre transform) by the rate function, which governs the behavior of the unbiased ensemble, and which, for Model 2, supports only one dynamical phase~\footnote{Note that neither of the the ``phases'' implied by the biased ensemble, $a = 0,1$ [inset \f{fig5}(b)], correspond to the stable phase $a_0=1/2$ of the model [\f{fig3}(b), \f{fig5}(c)].}.
\\~
\indent In such cases it is helpful, in order to establish a consistent physical picture, to use a method of rate-function reconstruction able to cope with non-convex functions\c{ioffe1993two,touchette2009large,ray2017importance,paga2017large,ferre2018adaptive}. In \f{fig7} we show rate functions computed using one such method, the reference-model method of Refs.\c{klymko2017rare,lattice2} (a form of nonequilibrium umbrella sampling). We also show the typical activity $a_s$ of the associated biased ensemble (distinct from the biased ensemble corresponding to the scaled cumulant-generating function). $a_s$ changes sharply with $s$ at the transition point of Model 1 [\f{fig7}(a)], and so both this method and the standard method (which detects a kink in the scaled cumulant-generating function and a jump in the activity of the biased ensemble) agree on the presence of a transition. By contrast, $a_s$ changes smoothly in the limit of slow dynamics of Model 2 [\f{fig7}(b)], where the dynamical large-deviation formalism shows sharp features [\f{fig5}(b)]. For this model, both methods are valid ways of reconstructing the rate function [\f{fig_supp}(a)]; the difference in their internal behavior results from the fact that the methods reconstruct affine portions of the rate function in different ways. In this case the smooth behavior of the reference-model method suggests that the rate function can be recovered analytically even in the limit of vanishing rate constants. This observation motivates the rescaling of \s{mod2}, and the identification of a more natural speed for the large-deviation principle. On this speed the rate function is strictly convex [\f{fig5}(c), \f{fig6}(a)] and its Legendre transform is analytic, and both methods now agree that there is only one stable dynamical phase. 

We end by noting that this idea, of searching for the most natural speed of the large-deviation principle for a given model and set of parameters, may provide insight into the nature of singularities seen using the scaled cumulant-generating function approach. For instance, Ref.\c{pietzonka2016extreme} shows that the time-averaged velocity $u$ of an active walker in continuous space obeys a large-deviation principle $\rho(u) \sim \e^{-h(u)t }$, where $t$ is time and the rate function
\beq
\label{a1}
h(u) = \max_\lambda [\lambda u - \alpha(\lambda)]
\eeq
is obtained from the quantity
\beq
\label{a2}
\alpha(\lambda) = \frac{1}{2} \left(\sqrt{9 \eta ^2+4 \eta  \lambda +4 \lambda
   ^2}-3 \eta +2 \lambda ^2\right),
\eeq
written here using the parameters of Fig. 1 (left column) of that reference. Here $\eta$ is the basic rate of switching between the two internal states of the walker. For small values of $\eta$ the rate function $h(u)$ becomes broad, indicating long-timescale intermittency and large fluctuations. When $\eta \to 0$, $\alpha \to \lambda^2 + |\lambda|$ possesses a discontinuous derivative at $\lambda=0$. As a result, the rate function computed by Legendre transform, \eqq{a1}, acquires a flat portion near its minimum: see \f{fig6}(b). 

We can examine this region in more detail by seeking a large-deviation principle $\rho(u) \sim \e^{-t_\infty h_\infty(u)}$, cast in terms of a speed $t_\infty \equiv t/\eta^{-1}$ containing the large timescale $\eta^{-1}$. Returning to \eq{a2}, replacing $\lambda \to \lambda \eta$, and performing a Maclaurin expansion in $\eta$ gives $\alpha(\eta \lambda) = \eta \alpha_\infty(\lambda) + {\cal O}(\eta^2)$, where
\beq
 \alpha_\infty(\lambda)=\frac{1}{2} \left(\sqrt{4 \lambda
   (\lambda +1)+9}-3\right)
\eeq
is analytic at $\lambda=0$. The rescaled version of \eqq{a1} is 
\bea
\label{a3}
h_\infty(u)=\lim_{\eta \to 0} h(u)/\eta &=& \max_\lambda [\lambda u - \alpha_\infty(\lambda)] \nonumber \\
&=& \frac{1}{2} \left(-2 \sqrt{2-2 u^2}-u+3\right), \hspace{1cm}
\eea
for $-1<u<1$. 

\eqq{a3}, plotted in \f{fig6}(b), shows that a large-deviation principle (with speed $t/\eta^{-1}$) exists on the interval $-1<u<1$ in the limit of slow dynamics, and that the associated rate function describes a single typical behavior [small fluctuations about which are Gaussian, i.e. $h_\infty(u) \approx \frac{27}{32} \left(u-\frac{1}{3}\right)^2$]. In this case, considering the large-deviation principle on more than one speed reveals the singularity of the model to be similar to that of Model 2.
\vspace{0.15cm}
\section{Conclusions}
\label{conc}

Singularities appear in the large-deviation formalism for a variety of reasons. In stochastic processes controlled by long-tailed probability distributions, singular features in moment-generating functions signal the breakdown of the large-deviation principle; such systems are described by non-normalizable probability distributions\c{aghion2017large,rebenshtok2014non}. Singularities in moment-generating functions also arise in models for which the the large-deviation principle applies. Here we have studied two models that give rise to singularities and possess distinct types of dynamical behavior, one analogous to an equilibrium phase transition and one not. In analyzing these models it was helpful to use different methods of reconstructing the large-deviation rate function, as well as to calculate the rate function on different speeds. It is possible that similar methods may provide additional insight into other models that induce singularities in the dynamical large-deviation formalism.
~
\begin{acknowledgments}
I am grateful to Katherine Klymko for assistance with \f{fig3}, and to Hugo Touchette for comments on the paper. This work was performed at the Molecular Foundry, Lawrence Berkeley National Laboratory, supported by the Office of Science, Office of Basic Energy Sciences, of the U.S. Department of Energy under Contract No. DE-AC02--05CH11231. 
\end{acknowledgments}

\appendix

 \section{Comparison of different methods of reconstructing the rate function helps to resolve the physics underlying the observed singularity}
 \label{app1}
 \begin{figure*}[] 
    \centering
    \includegraphics[width=\linewidth]{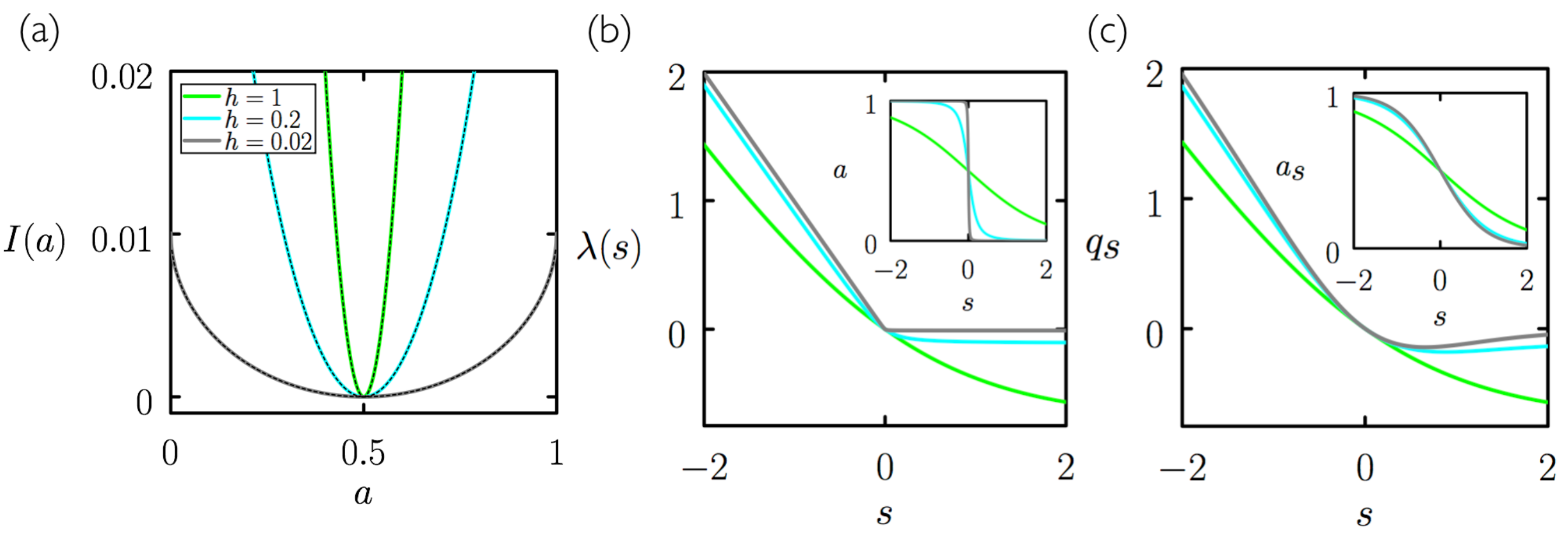} 
    \caption{(a) Large-deviation rate function $I(a)$ for Model 2 of \f{fig2} (parameters $\alpha=\beta=h/2$), computed using the dynamical large-deviation formalism (colored lines) and the reference-model method (black dashed lines). Panels (b) and (c) show the auxiliary quantities of the dynamical large-deviation approach and reference-model method, respectively.}
    \label{fig_supp}
 \end{figure*} 
In \f{fig_supp} we show rate functions for Model 2 of Fig. 2, calculated using the dynamical large-deviation formalism and the reference-model method of Refs.\c{klymko2017rare,lattice2} (a form of nonequilibrium umbrella sampling). This method makes use of a probability-conserving reference model whose transition probabilities are $p_{\rm ref}(i \to j)=p_s(i \to j)/\sum_j p_s(i \to j)$. The typical dynamics of the reference model can be used to recover the rare behavior of the original model. In the present case the rate function of the original model can be written
\beq
\label{rmm}
I(a_s) = -s a_s - q_s, \vspace{0.2cm}
\eeq
where the typical activity $a_s$ and typical path weight $q_s$ are computed using the equations given in Ref.\c{lattice2}. For the case $\alpha=\beta=h/2$ we have, in parametric form,
\beq
a_s=\frac{h(1-\e^s)-2}{(h-2) \e^{2 s}-2 h \e^s+h-2}
\eeq
and
\begin{widetext}
\bea
\label{rmm3}
q_s=\frac{\e^s \left((h-2) \e^s-h\right) \ln \left(h(\e^{-s}-1)/2+1\right)}{(h-2) \e^{2 s}-2
   h \e^s+h-2}-\frac{\left(h
   \left(\e^s-1\right)+2\right) \ln \left(\e^{-s}(1-h/2)+h/2\right)}{(h-2) \e^{2 s}-2
   h \e^s+h-2}.
\eea
\end{widetext}

The output of \eqq{rmm} is given by black dashed lines in panel (a) of \f{fig_supp}, and agrees with the rate function calculated by Legendre transform of the scaled cumulant-generating function. However, in the limit of small switching rate the auxiliary quantities $a_s$ and $q_s$ are distinct from the sharply-changing typical activity $a$ and scaled cumulant-generating function $\lambda$ of the dynamical large-deviation approach. This difference results from the fact that the two methods reconstruct linear portions of the rate function in different ways. Moreover, the lack of sharpness of the reference-model method suggests that the rate function in the limit of vanishing rate constants can be identified by choosing a new large-deviation speed [\s{mod2}]. On this new speed the rate function is strictly convex, and its Legendre transform is analytic.

%\bibliography{bib_v2}


\end{document}